# AI-assisted Protective Action: Study of ChatGPT as an Information Source for a Population Facing Climate Hazards

Xiangpeng Li[1,*], Yuqin Jiang[1], Ali Mostafavi[1]

[1] Ph.D student, UrbanResilience.AI Lab, Zachry Department of Civil and Environmental Engineering, Texas A&M University, College Station, TX, 77843; e-mail: xplli@tamu.edu

[1] Postdoctoral Researcher, Urban Resilience.AI Lab Zachry Department of Civil and Environmental Engineering, Texas A&M University, College Station, TX, 77843; e-mail: yuqinjiang@tamu.edu

[1] Associate Professor, Urban Resilience.AI Lab Zachry Department of Civil and Environmental Engineering, Texas A&M University, College Station, TX, 77843; e-mail: amostafavi@civil.tamu.edu

## Abstract

ChatGPT has been emerging as a novel information source, and it is likely that the public might seek information from ChatGPT while taking protective actions when facing climate hazards such as floods and hurricanes. Hence, it is critical to evaluate the reliability of the information provided by large language models, such as ChatGPT, about protective actions, such as obtaining hazard insurance, preparedness, evacuation, and adjustments to infrastructure disruptions, before and during climate hazard events. The objective of this study is to evaluate the accuracy and completeness of responses generated by ChatGPT when individuals seek information about aspects of taking protective actions. Multiple prompts related to types of protective actions (e.g., hazard insurance, preparedness, evacuation, and adjustments to service disruptions) were used to generate responses from ChatGPT. These responses were used in designing a web-based survey in which 38 emergency managers, with an average experience of 11 years, from organizations in the United States, evaluated the accuracy and completeness of the responses. The survey analysis results indicated that: (1) the emergency managers considered the responses provided by ChatGPT as accurate and complete to a great extent; (2) it was statistically verified in evaluations that the generated information was accurate, but lacked completeness, implying that the extent of information provided is accurate; and (3) information generated for prompts related to hazard insurance received the highest evaluation, whereas the information generated related to evacuation received the lowest. This last result implies that, for complex, context-specific protective actions (such as evacuation), the information was rated as less complete compared with other protective actions. Also, the results showed that the perception of respondents regarding the utility of AI-assistive technologies (such as ChatGPT) for emergency preparedness and response improved after taking the survey and evaluating the information generated by ChatGPT. The findings from this study provide empirical evaluation regarding the utility of AI-assistive technologies for improving public decision-making and protective actions in disasters.

# 1. Introduction

Proper preparedness prior to an upcoming climate hazard, such as a hurricane, can save lives and reduce hazard impacts. During preparedness and when making decisions about taking protective actions (such as evacuation and adjustment to service disruptions), residents actively seek information from various sources (Zhang et al., 2019) to enable better decisions and to take timely actions (Lindell & Perry, 2012; Lindell, 2018). Announcements from federal government agencies, such as the Federal Emergency Management Agency (FEMA), National Hurricane Center (NHC), local emergency management or agencies (Dow & Cutter, 1998; Zhang et al., 2007), and mass media, such as The Weather Channel. Opinions and information from friends and relatives also play important roles in the gathering of real-time hazard information (Burnside et al., 2007; Eisenman et al., 2007; Kyne et al., 2019; Lindell et al., 2005). Social media have become a dominant information source for the general public during the last decade. Since the emergence of social media, social media platforms have appeared to be one of the most widely used information sources for people facing hazard events (Zhang et al., 2019; Fan et al., 2020a; Rajput et al., 2020). Studies have found that residents use Twitter to seek information before, during, and after hurricanes. Social media have become important references for individuals to make decisions and to take protective actions (Fan, et al., 2020a; Fan et al., 2020b; Yang et al., 2019). The wide adoption and use of social media platforms for information seeking in hazard events show that people would extensively adopt and use new technologies that help them seek information while taking protective actions in facing disasters. The emergence of large language models (LLMs), such as ChatGPT, makes available a novel source of information. The user-friendly interface and fast response facilitate the use of AI-powered technology as an information source. ChatGPT and other AI-powered tools have been proved useful for decision-making in marketing, engineering, medical, transportation, and other fields (Abduljabbar et al., 2019; Duan et al., 2019; Holzinger et al., 2019; Huang & Rust, 2021; Szolovits, 2019). With the growing availability of LLM technologies, it is very likely that people will rely on these technologies to seek information in future hazard events. Thus, it is critical to evaluate the adequacy and accuracy of the information generated by this new technology in response to protective action-related prompts regarding weather hazard information.

This study examines the extent to which ChatGPT, and the large language processing models behind it, can provide suitable responses to inform residents to take protective actions in response to an upcoming climate hazard. Specifically, we asked ChatGPT questions regarding four types of protective actions: (1) flood insurance; (2) general pre-hurricane preparation; (3) evacuation; and (4) adjustment to infrastructure disruptions (e.g., power outages and disrupted access to critical facilities). We collected answers from ChatGPT and sent those answers to emergency management agencies. We asked professional emergency managers to evaluate the accuracy and completeness of each answer ChatGPT provided. To our knowledge, this is the first study that reacts to the increasing interest in using ChatGPT for hazard-related decision-making. This study sheds light to the usefulness of the state-of-the-art AI-powered technology as an information source to improve residents' protective actions in disasters. The findings of this study also can inform the federal, state, and local emergency management about the potential uses of LLM technologies for enhancing public warning and response to hazard events.

## 2. Data Collection

Using the Qualtrics web-based survey platform, we created survey questions to examine the perception of emergency managers regarding the accuracy and completeness of information generated by ChatGPT, the most dominant and well-known large language model. We first asked ChatGPT questions regarding natural hazard insurance, preparedness, evacuation, and adjustment to infrastructure disruption (Table 1). Then we collected responses generated by ChatGPT and listed those answers in the survey. These questions were phrased to replicate the manner in which a resident of the impacted area would ask ChatGPT in the imminent approach of a hazard. The survey had 18 prompts within the four categories mentioned above. The survey utilizes ChatGPT (February 13, 2023 version) to present the questions in sequence. (The Appendix contains detailed responses generated by ChatGPT.) We designed the survey to elicit from emergency managers evaluations of the accuracy and completeness of the responses generated by ChatGPT for each prompt. Each question requested a response from among a range, with 1 being very inaccurate through 5, being very accurate. Similarly, very incomplete was assigned a value of 1, and very complete, a value of 5. In addition, we asked questions regarding the perceptions of respondents regarding the usefulness of AI technologies in disaster preparedness and response to examine the extent to which respondents' perceptions about technology would influence their responses.

We deployed the survey to US emergency managers by email and social media platforms in early March 2023. Our survey was reviewed and approved by the Texas &M University Institutional Review Board. We collected 38 completed responses from individuals in several states who work in emergency management departments, with an average experience of 11 years in emergency-related positions.

Table 1. Categories of Prompts

| Category | Prompts |
|---|---|
| Natural Hazard Insurance | Q1. I am not living in a floodplain area, shall I buy flood insurance?<br>Q2. How can I know if my home insurance coverage is adequate for possible hurricane damage? |
| Natural Hazard Preparedness | Q3. My area is outside the path of an upcoming hurricane, will I be safe?<br>Q4. A storm with possible flooding is heading to our area. What should I do to prepare?<br>Q5. A storm with possible flooding is heading to our area. What is the likelihood of a flood occurring in my area, and how can I find out more information about the risks?<br>Q6. A storm with possible flooding is heading to our area. How can I reduce the risk of damage to my property in preparing for a flood?<br>Q7. A hurricane is heading to our area, what supplies should I have on hand in case of a hurricane?<br>Q8. A hurricane is heading to our area. How many days of power outages shall I prepare for? |
| Evacuation | Q9. A hurricane is heading to our area, shall I evacuate?<br>Q10. A storm with possible flooding is heading to our area, I have not received any evacuation order. What should I do?<br>Q11. A hurricane is heading to our area, and I have received an evacuation order. Where should I evacuate to?<br>Q12. A storm with possible flooding is heading to our area. What should I do if I cannot evacuate?<br>Q13. A storm with possible flooding is heading to our area tomorrow, is it late to evacuate? |
| Adjustments to Infrastructure Disruptions | Q14. Our area is affected by a hurricane and a power outage. How long can a power outage last during a major hurricane?<br>Q15. I rely on powered medical equipment and there is a power outage during a hurricane. What should I do?<br>Q16. Our area is affected by a hurricane and power outage, How can I stay informed about power outages during a hurricane?<br>Q17. There is a boil water notice in our area because of flooding. What should I do?<br>Q18. My area is affected by flooding and roads are inundated, What should I do if I need medical care |

## 3. Analysis and Results

### 3.1. Overall Accuracy and Completeness

In this section, we present the survey results. Figures 1 and 2 present the responses to the 18 completeness and accuracy survey questions. For example, there are two prompts in natural hazard insurance. In "Q1: I am not living in a floodplain area, shall I buy flood insurance?" Among the 38 responses, 23.68% of respondents found the ChatGPT-generated answers to be very accurate, 55.27% thought they were accurate, 13.16% deemed them somewhat accurate, 5.26% viewed them as inaccurate, and 2.6% considered them very inaccurate.

Accordingly, we can compare the proportion of each rating. The percentage of very accurate responses is significantly higher in the evaluation of ChatGPT responses related to the Natural Hazard Preparedness category than in the other categories. Furthermore, the assessment results for accuracy and completeness are overall favorable as the combined score of either very accurate and accurate or very complete and complete, surpasses the 50% threshold (indicating that more than half of the respondents rated ChatGPT responses as either very accurate or accurate and very complete or complete overall). This result shows the overall positive perception of emergency managers regarding the information generated by ChatGPT was accurate and complete.

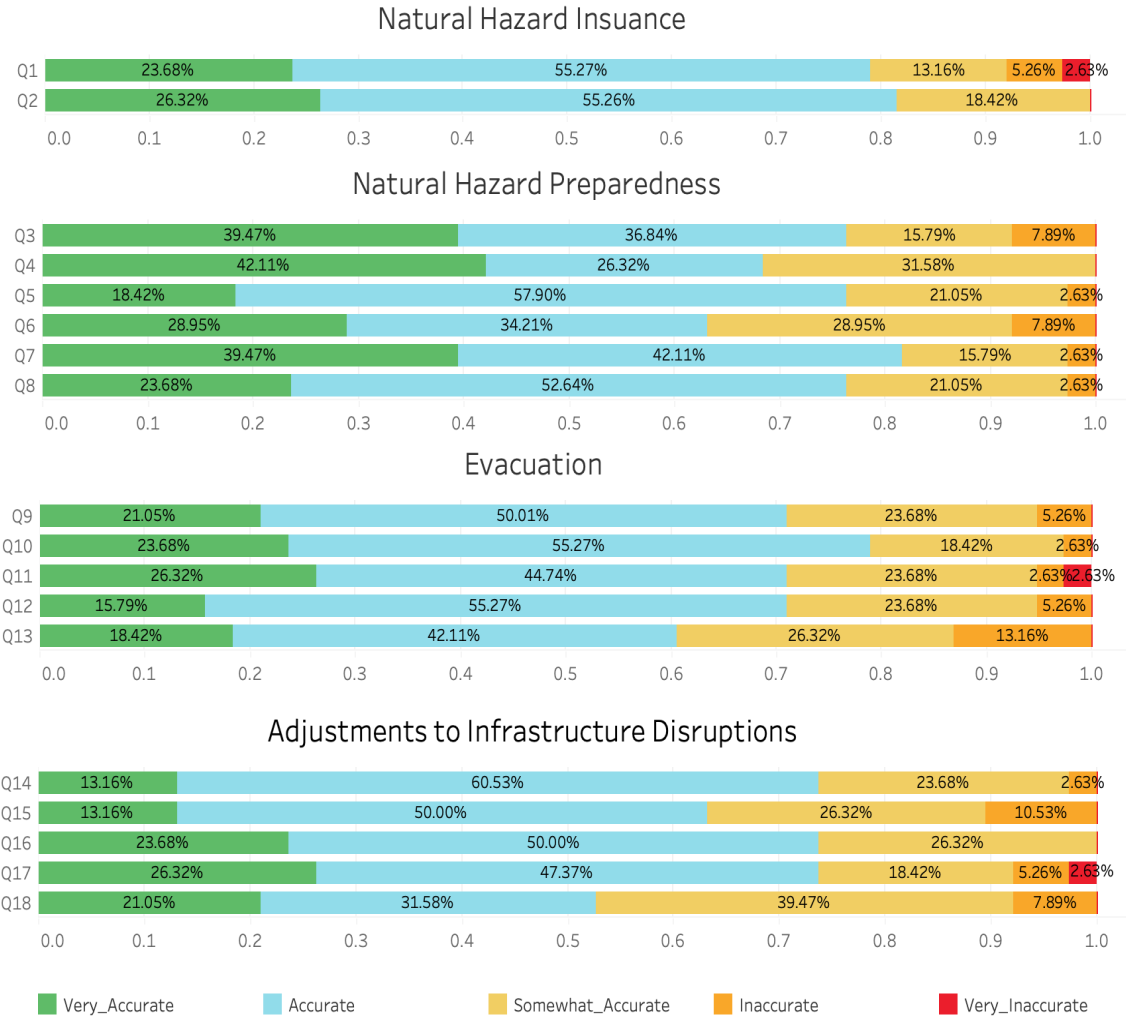

Figure 1. Perceived accuracy of ChatGPT responses: green corresponds to very accurate; blue, accurate; yellow, somewhat accurate; orange, inaccurate; and red, very inaccurate.

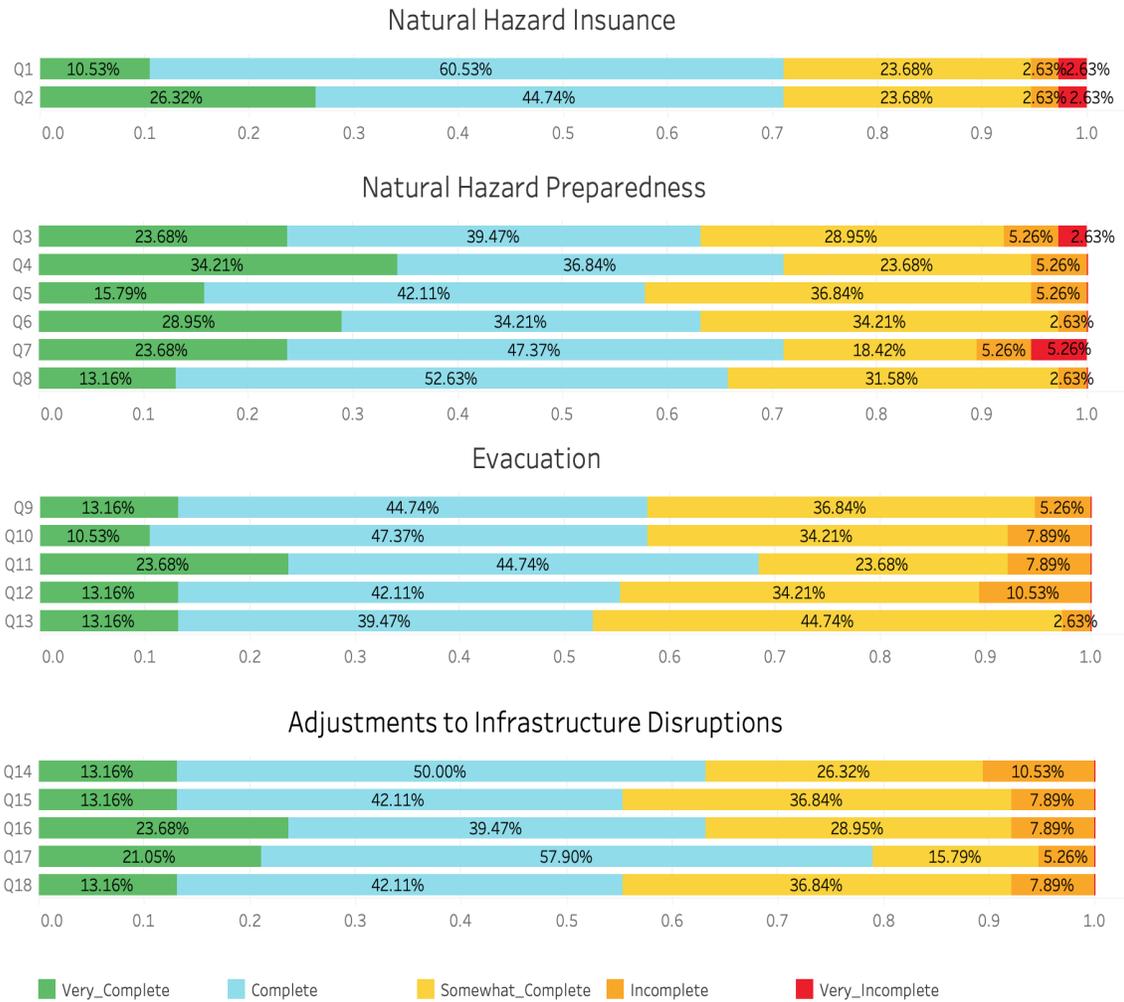

Figure 2. Perceived completeness of ChatGPT responses: green corresponds to very accurate, blue, accurate; yellow, somewhat accurate; orange, inaccurate; and red, very inaccurate.

## 3.2. Comparison of Evaluation on Each Category

We compared the responses across different categories of protective actions. Figures 3 and 4 illustrate the total accuracy, inaccuracy, completeness, and incompleteness for each category. For example, Figure 3 displays the total accuracy percentage (Sum_accurate), a summation of the percentages of very accurate and accurate, with a solid blue line. The mean value, 71.78%, of the Sum_accurate for each prompt is indicated by the horizontal dashed blue line. The total incorrect rate (Sum_Inaccurate) is depicted by a bold orange, while the average of the Sum_inaccurate, 5.11%, is indicated by the dashed orange horizontal line. Using these graphs, we can compare the total accuracy and inaccuracy of each category. For example, most data points that lie below mean accuracy or above mean inaccuracy are concentrated in the categories of evacuation and adjustments to infrastructure disruptions. Likewise, the categories of evacuation and adjustments to infrastructure disruptions also have a greater proportion of responses that fall below the mean completeness or above the mean incompleteness. This result suggests that the ChatGPT responses related to Evacuation and Adjustments to Infrastructure Disruptions are perceived to be less accurate and less complete compared with the responses related to the other two protective actions.

Tables 2 and 3 summarize the accuracy and completeness ratios derived from Figure 3 and Figure 4, respectively. For example, we examined the Sum_accuracy values that were greater than Mean_Accuracy, as well as Sum_Inaccurate values that were less than Mean_Inaccuracy when assessing accuracy. This analysis enabled us to compare the accuracy and inaccuracy percentages for each prompt. Table 2 shows that the proportion of Sum_Accurate values that exceeded Mean_Accuracy was 4 out of 6, or 2/3. Similarly, the proportion of Sum_Inaccurate values that were below Mean_Inaccuracy was 2/3, resulting in a product of 4/9. Accordingly, we ranked the assessment of ChatGPT responses for each category: Natural Hazard Insurance > Natural Hazard Preparedness > Adjustments to Infrastructure Disruptions > Evacuation. This result is intuitive since the information required for protective actions related to evacuation and infrastructure disruptions adjustments require context-specific details. The prompts we provided to ChatGPT were meant to be generic and did not have context-specific details.

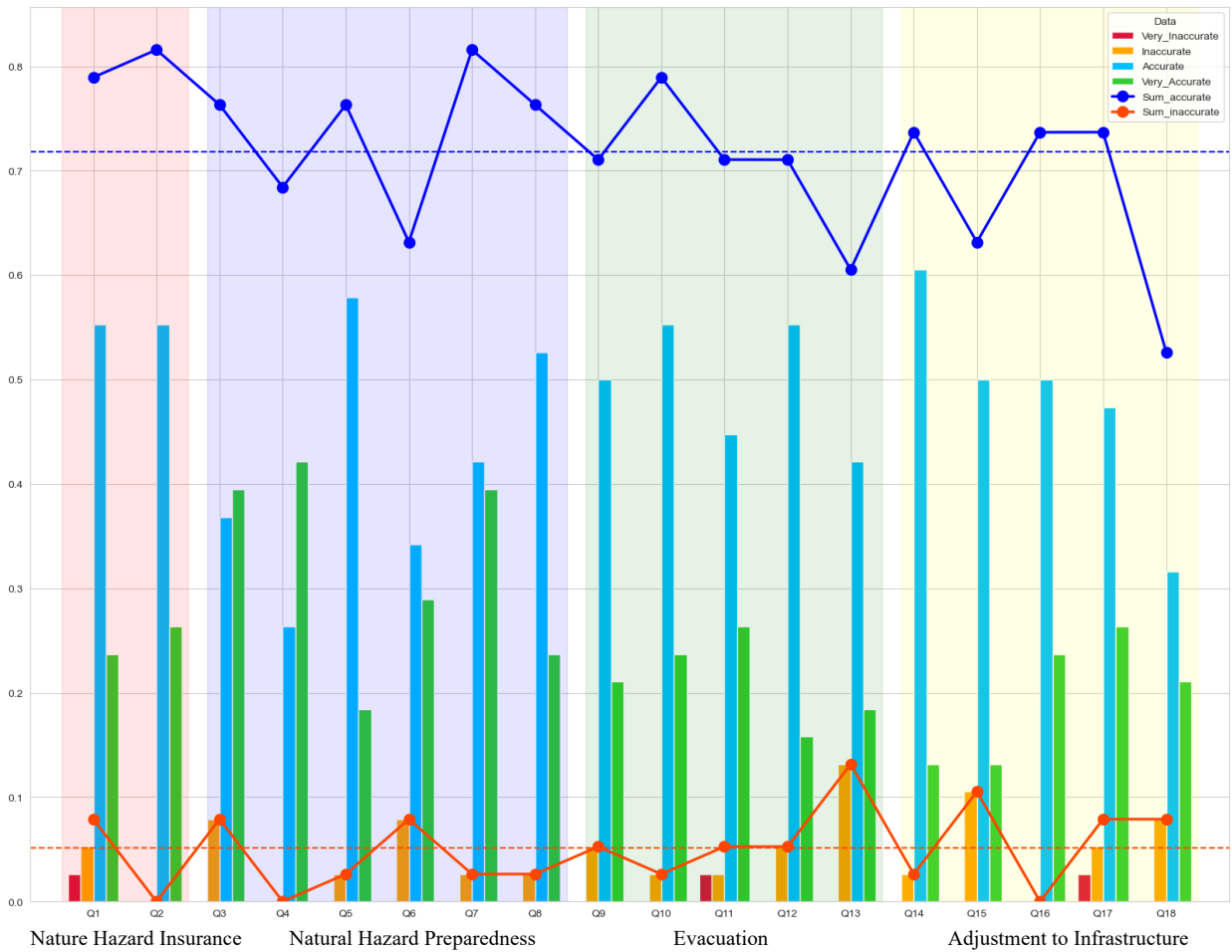

Figure 3. Overall perceived accuracy of responses for prompts in each protective action category.

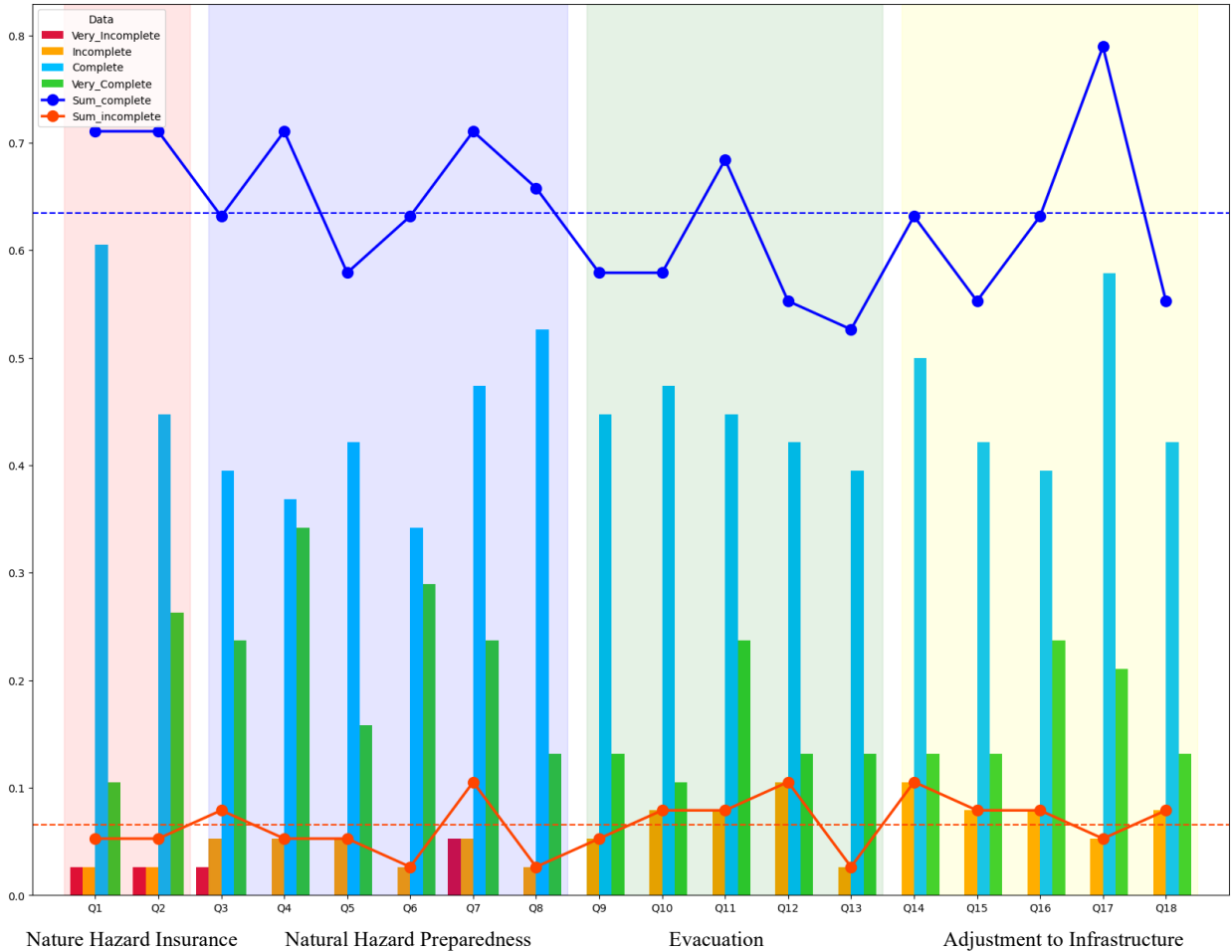

Figure 4. Figure 3. Overall perceived completeness of responses for prompts in each protective action category:

Table 2. Overall accuracy score of responses in each category

|  | Sum_Accuracy > Mean_Accuracy | Sum_Inaccuracy < Mean_Inaccuracy | Product |
|---|---|---|---|
| Insurance | 1 | 1/2 | 1/2 |
| Preparedness | 2/3 | 2/3 | 4/9 |
| Evacuation | 1/5 | 1/5 | 1/25 |
| Adjustment | 3/5 | 2/5 | 6/25 |

Table 3. Overall completeness score of responses in each category

|  | Sum_Complete > Mean_Complete | Sum_Incoplete < Mean_Incomplete | Product |
|---|---|---|---|
| Insurance | 1 | 1 | 1 |
| Preparedness | 5/6 | 2/3 | 5/9 |
| Evacuation | 1/5 | 4/5 | 4/25 |
| Adjustment | 3/5 | 2/5 | 6/25 |

## 3.3. Comparison of Accuracy and Completeness Ratings

The bar chart in Figure 5 illustrates the average evaluations of accuracy and completeness for each prompt. The accuracy and completeness scores are both between 3.5 and 4, indicating that the responses were evaluated by emergency managers to be quasi-accurate and quasi-complete. The blue dashed line illustrates the mean accuracy of 3.91; the yellow dashed line reveals mean completeness of 3.74. The average accuracy is higher than the average completeness, supported by the area chart shown in Figure 6. To compare the accuracy and completeness ratings, we performed a t-test and obtained a t-value of 3.44 and a highly significant p-value of 0.0016. The results suggest a significant difference between the means of accuracy and completeness. That is, the respondents rated ChatGPT responses higher in terms of accuracy compared with completeness. In other words, the information provided by ChatGPT might not be fully complete, but the extent of the information provided is accurate.

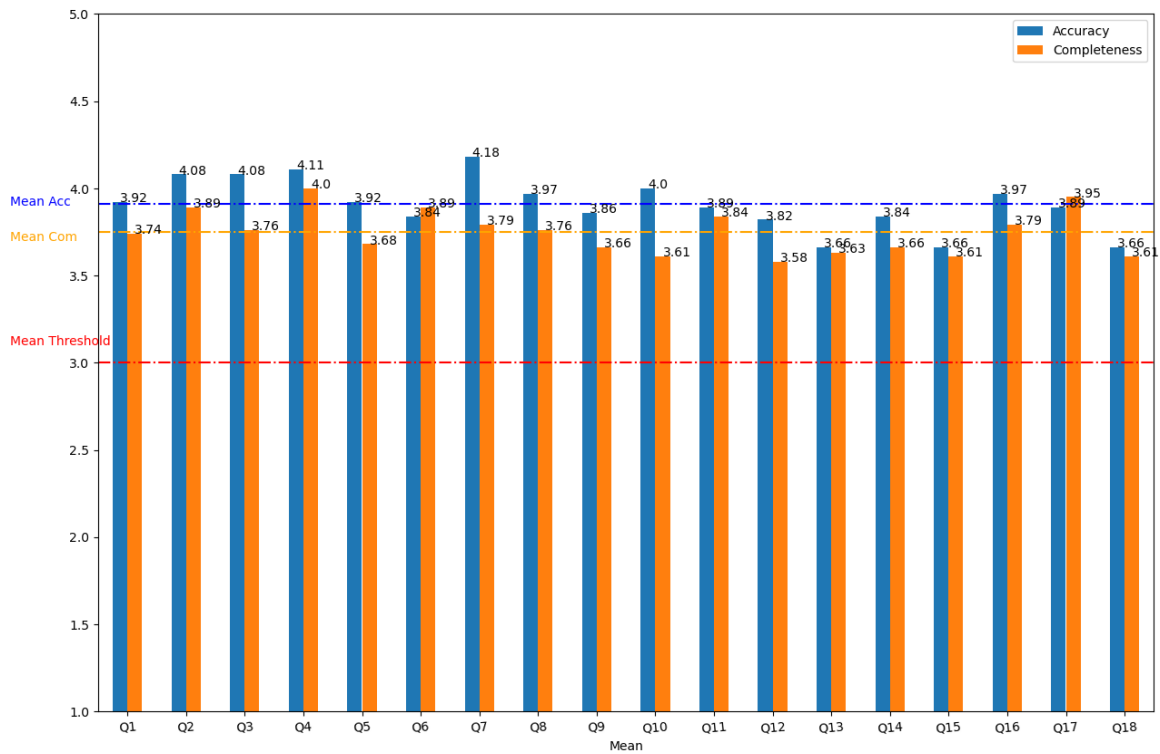

Figure 5. The mean value of accuracy and completeness ratings across different questions

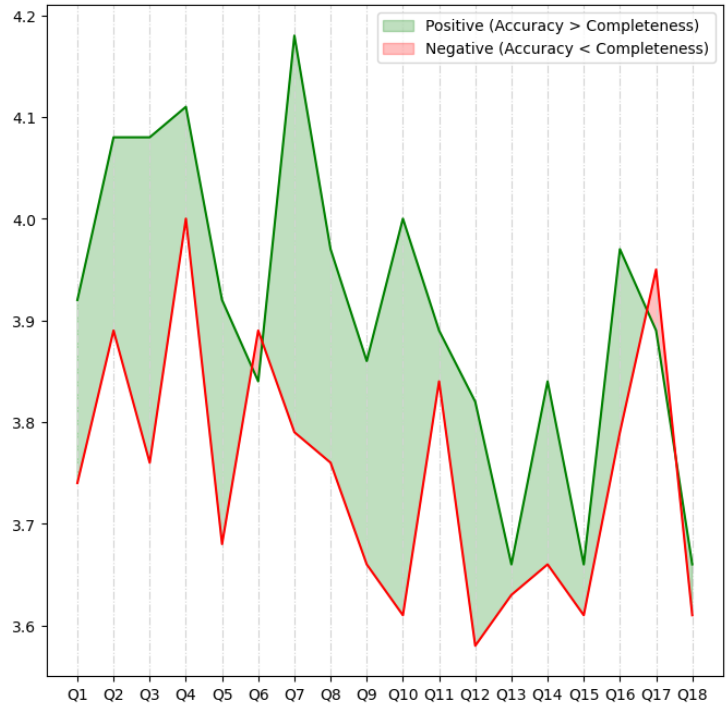

Figure 6. Area Chart of Accuracy and Completeness

## 3.4. Respondents' Perception of Technology

Before participating, 45% of the respondents had no prior experience with ChatGPT or were unfamiliar with ChatGPT. At the beginning and the end of the survey, we asked the participants to rate the potential of artificial intelligence technologies, like ChatGPT, to improve the public's ability to respond to natural disasters on a scale of 1 to 5. Before taking the survey, the overall average score was 3.54, which is higher than the theoretical average value of 3. After completing the survey, the respondents' average rating for this question increased to 3.84. This result suggests that the perception of emergency management professionals regarding the utility of AI technologies such as ChatGPT improved after the review and evaluation of the information generated by ChatGPT. According to a few remarks provided by the respondents, some respondents thought that it is essential to recognize that an automated tool, with the current level of technology, cannot offer a simple one- to two-paragraph answer on complex subjects, such as whether to evacuate in the event of a potential flood. In the same way, such a tool cannot provide advice on how to safely evacuate through rush hour traffic in Los Angeles. On the other hand, some respondents expressed initial doubts about the technology's reliability before participating. Following their participation, however, they felt considerably more assured that the technology can offer valuable guidance for emergency preparedness and response to local residents. Some respondents appreciated that ChatGPT encourages users to reach out to their local emergency management or other authorities.

## 4. Concluding Remarks

This study is the first attempt to examine the potential of AI technologies, such as large language models, as an emerging information source to improve the public's protective actions when facing hazard events. The findings of this study show that emergency management professionals highly rated the accuracy and completeness of ChatGPT s responses, suggesting a positive perception of LLM technologies for providing useful information to inform people's protective actions. Essentially, the findings show that emergency management professionals acknowledged the value of ChatGPT in assisting people in seeking information about natural hazard insurance, preparedness, evacuation, and adjustment to infrastructure disruptions. Among the four categories of pre-hazard-related questions, natural hazard insurance received the highest evaluation. Insurance-related queries are more generic and require fewer context-specific details, thus ChatGPT is better able to deliver accurate and complete responses. For insurance-related questions, ChatGPT can provide clear information for people to check related official websites, which cover the most relevant insurance-related content about their hazard insurance. On the other hand, ChatGPT responses related to evacuation received the lowest evaluation scores. Evacuation is a complicated decision that is influenced by multiple factors, such as car ownership, the availability of finding lodging, accommodation for pets, family and friends' decisions, and work requirements (Huang, et al., 2012; 2016; Lindell et al., 22005; Dow & Cutter, 1998; Jiang et al., 2021). Therefore, generic evacuation suggestions can hardly address each individual's personal information needs for decision-making. Evacuation decisions require more tailored information and are usually context specific. In addition, in several comments received after the survey, respondents also remarked that ChatGPT's responses to evacuation questions lacked detail or were not customized to their particular circumstances. Accurate and timely information is essential for impacted people to make correct decisions. The application of AI-powered tools in making more personal decisions to address different demands requires further study.

The evaluation of responses also revealed that ChatGPT could formulate accurate information in response to different protective action-related prompts. This study found that the score for completeness is lower than that for accuracy, meaning that ChatGPT may not always provide a comprehensive response or in-depth clarification of the subject, though the information provided is accurate. A sampling of feedback from the survey respondents included comments such as "the answer can be more comprehensive," "the answer is generic," or "it is not specific to the situation." These comments pointed to the fact that ChatGPT cannot handle specific context-specific cases. This limitation can also be due to the generic nature of the prompts we used in this study. Future studies could examine more hazard-specific and context-specific prompts in the evaluation of ChatGPT responses. Overall, despite these limitations related to the completeness of the answers provided by the ChatGPT, its accuracy is still an asset that can be relied upon for specific tasks and applications.

Based on the survey results, emergency management professionals endorsed that ChatGPT, as one of the most representative AI-powered interactive LLMs, is able to provide useful information to the public regarding an upcoming natural hazard. The usefulness of ChatGPT is recognized and acknowledged. Currently, the main concerns are the comprehensiveness and completeness in the answers ChatGPT provided. Specifically, ChatGPT's answers are generic and fail to address customized demands.

This study is the first step toward evaluating AI-assisted protective actions using technologies such as LLM. The potential opportunities to utilize LLM technologies for emergency response and disaster resilience are enormous and exciting. One of the primary benefits of LLM is its ability to process large amounts of data and respond to user inquiries in real time. During a natural disaster, there may be a flood of information from various sources, including government agencies, news outlets, and social media platforms. LLMs can help sift through this information to provide users with up-to-date information on evacuation routes, shelter locations, and other critical information that can help individuals make informed decisions and stay safe during an emergency. Another benefit of LLMs is their ability to communicate in multiple languages. During a natural disaster, it is not uncommon for individuals who do not speak the primary language of the affected region to be impacted. LLMs can help bridge this communication gap by providing information in a variety of languages, making information more accessible and enabling all affected to understand and respond appropriately to the situation. In addition to providing information, LLMs can also assist emergency responders in coordinating their efforts. For example, LLMs can help first responders quickly identify areas of high need and prioritize their resources accordingly. It can also help emergency responders communicate with one another more efficiently, allowing for better coordination and faster response times. Perhaps most importantly, LLMs can help provide emotional support to those affected by natural disasters. During times of crisis, individuals may feel overwhelmed, anxious, and scared. LLM can provide a sympathetic ear and offer suggestions for coping with these emotions. This can be particularly helpful for individuals who may not have access to mental health services during an emergency. Of course, there are some limitations to the use of LLMs in emergency response. For example, the accuracy of the information provided by LLMs relies heavily on the quality of the data it is trained on. If the data is outdated or inaccurate, LLM may provide incorrect information. Additionally, not all individuals may have access to the internet or other means of communicating with LLM during an emergency. Future studies should evaluate these opportunities and limitations to further realize the full potential of AI-assisted emergency response to climate hazards.

## Acknowledgment

We would like to express our most sincere appreciation to Ashely Morris for generously volunteering her time and expertise to help with the data collection process for this research project. Her support and enthusiasm for this project were invaluable and greatly appreciated.

## Declaration of interests
The authors declare that they have no known competing interests.

# References:

Abduljabbar, R., Dia, H., Liyanage, S., & Bagloee, S. A. (2019). Applications of artificial intelligence in transport: An overview. *Sustainability*, *11*(1), 189.

Burnside, R., Miller, D. S., & Rivera, J. D. (2007). The impact of information and risk perception on the hurricane evacuation decision-making of greater New Orleans residents. *Sociological Spectrum*, *27*(6), 727–740.

Dow, K., & Cutter, S. L. (1998). *Crying wolf: Repeat responses to hurricane evacuation orders*.

Duan, Y., Edwards, J. S., & Dwivedi, Y. K. (2019). Artificial intelligence for decision making in the era of Big Data–evolution, challenges and research agenda. *International Journal of Information Management*, *48*, 63–71.

Eisenman, D. P., Cordasco, K. M., Asch, S., Golden, J. F., & Glik, D. (2007). Disaster planning and risk communication with vulnerable communities: Lessons from Hurricane Katrina. *American Journal of Public Health*, *97*(Supplement_1), S109–S115.

Fan, C., Jiang, Y., & Mostafavi, A. (2020a). Social sensing in disaster city digital twin: Integrated textual–visual–geo framework for situational awareness during built environment disruptions. Journal of Management in Engineering, 36(3), 04020002.

Fan, C., Jiang, Y., Yang, Y., Zhang, C., & Mostafavi, A. (2020b). Crowd or hubs: information diffusion patterns in online social networks in disasters. International journal of disaster risk reduction, 46, 101498.

Freberg, K., Saling, K., Vidoloff, K. G., & Eosco, G. (2013). Using value modeling to evaluate social media messages: The case of Hurricane Irene. *Public Relations Review*, *39*(3), 185–192.

Holzinger, A., Langs, G., Denk, H., Zatloukal, K., & Müller, H. (2019). Causability and explainability of artificial intelligence in medicine. *Wiley Interdisciplinary Reviews: Data Mining and Knowledge Discovery*, *9*(4), e1312.

Huang, S. K., Lindell, M. K., Prater, C. S., Wu, H. C., & Siebeneck, L. K. (2012). Household evacuation decision making in response to Hurricane Ike. Natural Hazards Review, 13(4), 283-296.

Huang, S., M. K. Lindell, and C. S. Prater. 2016. Who leaves and who stays? A review and statistical meta- analysis of hurricane evacuation studies. Environment and Behavior 48 (8):991–1029.

Huang, M.-H., & Rust, R. T. (2021). A strategic framework for artificial intelligence in marketing. *Journal of the Academy of Marketing Science*, *49*, 30–50.

Jiang, Y., Li, Z., & Cutter, S. L. (2021). Social distance integrated gravity model for evacuation destination choice. International Journal of Digital Earth, 14(8), 1004-1018.

Kapucu, N., Berman, E. M., & Wang, X. H. (2008). Emergency information management and public disaster preparedness: Lessons from the 2004 Florida hurricane season. *International Journal of Mass Emergencies & Disasters*, *26*(3), 169–197.

Kim, J., Bae, J., & Hastak, M. (2018). Emergency information diffusion on online social media during storm Cindy in US. *International Journal of Information Management*, *40*, 153–165.

Kyne, D., Donner, W., & Lomeli, A. S. (2019). Information seeking and evacuation decision making in hypothetical hurricane events in the Rio Grande Valley, Texas. *International Journal of Emergency Management*, *15*(1), 54–80.

Lindell, M. K., Lu, J.-C., & Prater, C. S. (2005). Household decision making and evacuation in response to Hurricane Lili. *Natural Hazards Review*, *6*(4), 171–179.

Lindell, M. K., & Perry, R. W. (2012). The protective action decision model: Theoretical modifications and additional evidence. Risk Analysis: An International Journal, 32(4), 616-632.

Lindell, M. K. (2018). Communicating imminent risk. Handbook of disaster research, 449-477.

Rajput, A. A., Li, Q., Zhang, C., & Mostafavi, A. (2020). Temporal network analysis of inter-organizational communications on social media during disasters: A study of Hurricane Harvey in Houston. International journal of disaster risk reduction, 46, 101622.

Szolovits, P. (2019). *Artificial intelligence in medicine*. Routledge.

Yang, Y., Zhang, C., Fan, C., Yao, W., Huang, R., & Mostafavi, A. (2019). Exploring the emergence of influential users on social media during natural disasters. International Journal of Disaster Risk Reduction, 38, 101204.

Zhang, F., Morss, R. E., Sippel, J., Beckman, T., Clements, N., Hampshire, N., Harvey, J., Hernandez, J., Morgan, Z., & Mosier, R. (2007). An in-person survey investigating public perceptions of and responses to Hurricane Rita forecasts along the Texas coast. *Weather and Forecasting*, *22*(6), 1177–1190.

Zhang, C., Fan, C., Yao, W., Hu, X., & Mostafavi, A. (2019). Social media for intelligent public information and warning in disasters: An interdisciplinary review. International Journal of Information Management, 49, 190-207.

# Appendix

## 1. Natural Hazard Insurance

**Q1. I am not living in a floodplain area, shall I buy flood insurance?**

The response generated by ChatGPT:

It's always a good idea to consider purchasing flood insurance even if you don t live in a designated floodplain area. While floodplains are considered higher-risk areas for flooding, it's important to remember that floods can occur anywhere, and even a small amount of water damage can be costly to repair.

Additionally, it's worth noting that standard homeowners or renters' insurance policies typically don t cover damage from flooding. So, if you're concerned about the financial risk of flood damage to your property or belongings, you may want to consider purchasing a separate flood insurance policy.

You can find out more about flood insurance options by speaking with a licensed insurance agent, or by visiting the National Flood Insurance Program website at http://www.floodsmart.gov.

**Q2. How can I know if my home insurance coverage is adequate for possible hurricane damage?**

The response generated by ChatGPT:

1. Check your coverage limits: Make sure that your coverage limits for your dwelling, personal property, and additional living expenses are high enough to cover the potential cost of repairing or rebuilding your home after a hurricane. Remember to take into account the cost of materials and labor in your area, which can vary widely.

2. Check your deductible: Make sure you know what your hurricane deductible is and that you're comfortable with the amount. Hurricane deductibles are often higher than regular deductibles, and they're typically a percentage of your home's insured value.

3. Check for exclusions: Review your policy for any exclusions or limitations related to hurricane damage. For example, some policies may exclude flood damage, or may have limited coverage for wind damage.

4. Check for additional coverage: Consider purchasing additional coverage for things like sewer backup, debris removal, or mold remediation, which may be more likely to occur after a hurricane.

5. Consider purchasing flood insurance: Remember that standard homeowners or renters insurance policies typically do not cover flood damage. If you live in a coastal area, or an area that is prone to flooding, you may want to consider purchasing a separate flood insurance policy.

If you have any questions or concerns about your insurance coverage, it's always a good idea to speak with your insurance agent to make sure you have the coverage you need.

## 2. Natural Hazard Preparedness

**Q3. My area is outside the path of an upcoming hurricane, w ill I be safe?**

The response generated by ChatGPT:

While being outside the path of a hurricane can certainly reduce the risk of direct impact from the storm, it's important to keep in mind that hurricanes can still cause significant damage even outside of their direct path.

Hurricanes can bring strong winds, heavy rain, storm surge, and even tornadoes that can extend well beyond the eye of the storm. The strength and extent of these impacts will depend on a variety of factors, including the size and strength of the storm, the proximity to your location, and the terrain and infrastructure in your area.

It's always a good idea to stay informed about the track and potential impacts of any upcoming storms, even if they're not expected to make a direct hit on your area. You can do this by monitoring local news and weather updates, signing up for emergency alerts from your local government, and having a plan in place in case you need to evacuate or take other protective measures.

If you're concerned about the potential impact of a hurricane on your property, you may want to consider reviewing your insurance coverage and taking steps to prepare your home and belongings for a storm. This may include things like trimming trees and securing loose outdoor items, as well as reviewing your insurance policy to make sure you have adequate coverage for potential storm damage.

**Q4. A storm with possible flooding is heading to our area, what should I do to prepare?**

The response generated by ChatGPT:

1. Stay informed: Keep track of weather updates and follow any advice or orders from local authorities. Sign up for emergency alerts and make sure you know the evacuation routes in your area.

2. Secure your property: Move any valuable items to higher ground or to an upper level of your home. If you have time, sandbag the perimeter of your property to help prevent water from entering your home.

3. Stock up on supplies: Make sure you have enough food, water, medications, and other necessary supplies to last at least a few days. Keep a battery-powered radio and extra batteries on hand in case of power outages.

4. Protect your home: Check your gutters and downspouts to make sure they're clear and functioning properly. Turn off your electricity and gas supply if you need to evacuate. If you have time, consider installing storm shutters or boarding up your windows to help protect them from debris.

5. Evacuate if necessary: If local authorities recommend or order an evacuation, follow their instructions and leave as soon as possible. Don t wait until the last minute, as flood waters can rise quickly and become dangerous.

Remember that flooding can be unpredictable, and it's important to prioritize your safety above all else. Take the necessary precautions to protect yourself, your family, and your property, and stay informed throughout the storm.

**Q5. A storm with possible flooding is heading to our area, what is the likelihood of a flood occurring in my area, and how can I find out more information about the risks?**

The response generated by ChatGPT:

The likelihood of flooding occurring in your area during a storm will depend on a variety of factors, including the intensity of the storm, the amount and duration of rainfall, and the local topography and infrastructure. However, even areas that are not typically prone to flooding can experience flash flooding during heavy rainfall or storm surge.

To find out more information about the risks of flooding in your area, you can do the following:

1. Check flood maps: FEMA provides online flood maps that can give you an idea of your area s flood risk. You can enter your address to view the flood zone and Base Flood Elevation (BFE) for your property.

2. Contact your local emergency management office: Your local emergency management office can provide information about your area s flood risk and any evacuation plans. They may also have information about flood insurance options.

3. Monitor weather reports: Stay informed about the weather forecast for your area and any potential risks of flooding. Local news outlets and the National Weather Service can provide updates on the storm's expected path and severity.

4. Consult with a licensed insurance agent: An insurance agent can help you understand your insurance coverage and any options for additional coverage, such as flood insurance.

It's important to take the necessary precautions to protect yourself and your property from potential flooding, even if the risk is relatively low. Be prepared with emergency supplies, develop an evacuation plan, and stay informed throughout the storm.

**Q6. A storm with possible flooding is heading to our area, how can I reduce the risk of damage to my property in preparing for a flood?**

The response generated by ChatGPT:

There are several steps you can take to reduce the risk of damage to your property in preparing for a flood:

1. Clear gutters and downspouts: Make sure your gutters and downspouts are clear of debris and functioning properly to help water flow away from your property.

2. Elevate or relocate valuables: Move any valuable items to a higher floor or to an upper level of your home to help protect them from flood waters. Consider storing items that can't be elevated, such as electronics or important documents, in waterproof containers.

3. Install a sump pump: A sump pump can help remove water from your basement or lower level in the event of flooding.

4. Seal cracks and openings: Seal any cracks or openings in your foundation or walls to help prevent water from entering your property.

5. Consider flood barriers: There are various types of flood barriers available that can help prevent water from entering your property. You may want to consider installing these if you live in a high-risk flood area.

6. Turn off utilities: Turn off your electricity and gas supply if you need to evacuate, to reduce the risk of fire and other hazards.

7. Have an emergency kit ready: Prepare an emergency kit with food, water, and other necessary supplies to last at least a few days in case of flooding or evacuation.

Remember that floodwaters can rise quickly and become dangerous, so your safety should always be your top priority. If local authorities recommend or order an evacuation, follow their instructions and leave as soon as possible.

**Question 7. A hurricane is heading to our area, what supplies should I have on hand in case of a hurricane?**

The response generated by ChatGPT:

If a hurricane is heading to your area, it's important to be prepared with emergency supplies in case of power outages, flooding, or other damage. Some important supplies to have on hand include:

1. Water: Store at least one gallon of water per person per day for at least three days. It's also a good idea to have some water purification tablets or a filtration system in case you need to purify additional water.

2. Non-perishable food: Have a supply of non-perishable food that doesn't require refrigeration or cooking, such as canned goods, energy bars, or dried fruit.

3. First aid kit: A first aid kit should include items such as bandages, antiseptic, pain relievers, and any necessary prescription medications.

4. Flashlights and batteries: Make sure you have several flashlights and extra batteries on hand in case of power outages.

5. Personal hygiene items: Have supplies such as toilet paper, hand sanitizer, and wet wipes in case of water or power outages.

6. Cash: Keep some cash on hand in case ATMs are unavailable or electronic payment systems

are down.

7. Important documents: Keep important documents such as identification, insurance policies, and medical records in a waterproof container.8. Tools and supplies: Have basic tools such as a wrench, pliers, and a screwdriver, as well as duct tape and plastic sheeting to help secure your home.

9. Pet supplies: If you have pets, make sure you have enough food, water, and other supplies to last at least a few days.

Remember to keep your emergency supplies in a portable container or bag that you can easily grab in case of an evacuation. Stay informed about the hurricane and any necessary preparations or evacuations, and follow the instructions of local authorities to ensure your safety.

**Question 8. A hurricane is heading to our area, how many days of power outage shall I prepare for?**

The response generated by ChatGPT:

It's difficult to predict exactly how long a power outage might last during or after a hurricane, as it will depend on the severity of the storm, the extent of the damage, and the availability of repair crews. However, it's generally a good idea to be prepared for a power outage lasting at least a few days.

The Federal Emergency Management Agency (FEMA) recommends that individuals and households prepare for a power outage lasting for at least 72 hours, or three days. This means having enough food, water, and other necessary supplies on hand to last for this amount of time.

It's also important to be aware that power outages can sometimes last longer than expected, especially in the aftermath of a major hurricane. If you live in an area that is particularly prone to hurricane damage or power outages, you may want to consider preparing for a longer period of time, such as one week or more.

Regardless of the expected duration of the power outage, it's always a good idea to be prepared with plenty of emergency supplies to ensure your safety and comfort during and after the storm.

3. Evacuation

**Question 9. A hurricane is heading to our area, shall I evacuate?**

The response generated by ChatGPT:

Whether or not you should evacuate when a hurricane is heading to your area depends on several factors, including the severity of the storm, the risk of flooding, the location and type of your home, and any guidance or orders from local authorities.

If you live in a low-lying area or an area at high risk for flooding, you may be advised or required to evacuate even if your home is not directly in the path of the hurricane. Similarly, if you live in a mobile home or other structure that may not be able to withstand high winds or storm surges, you may also be advised to evacuate.

In general, it's a good idea to stay informed about the latest weather forecasts and updates from local authorities, including any evacuation orders or recommended preparations. If you are advised or ordered to evacuate, it's important to follow these instructions as soon as possible to ensure your safety.

If you are not advised to evacuate, but you are concerned about your safety or the safety of your home, you may still want to take precautions such as securing windows and doors, moving valuables to higher ground, and stocking up on emergency supplies.

Remember that hurricanes can be unpredictable and dangerous, so it's always best to err on the side of caution when it comes to your safety and the safety of your loved ones. If you have any doubts or concerns, you may want to speak with a local emergency management agency or other authority to get more information about your specific situation.

**Question 10. A storm with possible flooding is heading to our area. I have not received any evacuation order, what should I do?**

The response generated by ChatGPT:

If you have not received an evacuation order, but a storm with possible flooding is heading to your area, there are still several steps you can take to prepare and protect yourself and your property:

1. Stay informed: Monitor the latest weather forecasts and updates from local authorities to stay informed about the storm s progress and any potential risks.

2. Prepare your home: Secure windows and doors, move outdoor furniture and other loose items indoors, and clear gutters and drains to help prevent water damage.

3. Stock up on emergency supplies: Have enough food, water, medications, and other necessary supplies to last for several days, in case of power outages or other disruptions.

4. Stay indoors during the storm: Avoid driving or walking through flooded areas, and stay inside until the storm has passed.

5. Be ready to evacuate: Have a plan in place in case an evacuation order is issued, including knowing your evacuation route and having a "go bag" ready with important documents, medications, and other necessary items.

It's important to take any storm warnings or alerts seriously, and to be prepared for the possibility of flooding or other damage. Even if you have not received an evacuation order, it's always a good idea to take steps to protect yourself, your loved ones, and your property.

**Question 11. A hurricane is heading to our area, and I have received an evacuation order, where should I evacuate to?**

The response generated by ChatGPT:

If you have received an evacuation order due to an approaching hurricane, it's important to follow the guidance of local authorities and evacuate as soon as possible. The specific location to which you should evacuate will depend on a number of factors, including the severity of the storm and the availability of nearby shelters or other safe locations.

Here are some steps you can take to find a safe place to evacuate:

1. Check with local authorities: Local emergency management agencies and other authorities may provide information about available evacuation shelters or other safe locations. Check their websites or social media pages for updates and instructions.

2. Consider family and friends: If you have family or friends outside the affected area, you may be able to stay with them during the storm. Be sure to contact them in advance to make arrangements.

3. Look for hotels or other accommodations: If you are able to evacuate by car, you may be able to find a hotel or other temporary lodging outside the affected area. Be sure to make reservations in advance, as these may fill up quickly.

4. Use a mobile app: There are several mobile apps, such as the Red Cross Emergency App, that can provide information about available shelters and other resources during a disaster.

Remember to bring important documents, such as identification, insurance information, and medical records, with you when you evacuate. Also, be sure to follow any other instructions or guidance provided by local authorities, such as evacuation routes or transportation options.

**Question 12. A storm with possible flooding is heading to our area, what should I do if I cannot evacuate?**

The response generated by ChatGPT:

If a storm with possible flooding is heading to your area and you are unable to evacuate, there are several steps you can take to protect yourself and your property:

1. Stay indoors: If possible, stay inside your home or other secure location until the storm has passed.

2. Move to higher ground: If you are in an area that is prone to flooding, move to higher ground to reduce the risk of flood damage.

3. Turn off electricity and gas: If flooding is likely, turn off the electricity and gas in your home to reduce the risk of fire or electrical shock.

4. Secure your property: Move valuable items to higher ground and secure loose items that could be swept away by floodwaters.

5. Stock up on emergency supplies: Have enough food, water, medications, and other necessary supplies to last for several days, in case of power outages or other disruptions.

6. Stay informed: Monitor the latest weather forecasts and updates from local authorities to stay informed about the storm s progress and any potential risks.

7. Call for help: If you are in danger or need assistance, call 911 or your local emergency services for help.

It's important to take any storm warnings or alerts seriously and to be prepared for the possibility of flooding or other damage. If you cannot evacuate, taking these steps can help to protect yourself and your property as much as possible.

**Question 13. A storm with possible flooding is heading to our area tomorrow, is it late to evacuate?**

The response generated by ChatGPT:

If a storm with possible flooding is heading to your area and you are considering evacuating, it's always best to do so as early as possible. However, even if the storm is approaching quickly, it may not be too late to evacuate, especially if you are able to move quickly and follow all safety protocols.

If you are unsure whether it is too late to evacuate, you can check with local emergency management agencies or other authorities for guidance. They may be able to provide information about evacuation routes, shelters, and other resources that can help you make the decision.

In any case, it's important to prioritize your safety and the safety of your family. If you feel that you are in danger, or if local authorities have issued an evacuation order, you should move to a safe location as soon as possible. Remember to bring important documents, such as identification, insurance information, and medical records, with you when you evacuate.

If you are unable to evacuate, be sure to take other steps to protect yourself and your property, such as moving to higher ground, securing loose items, and stocking up on emergency supplies. Stay informed about the storm s progress and any potential risks, and be prepared to call for help if needed.

1. Adjustments to Infrastructure Disruptions.

   **Question 14. Our area is affected by a hurricane and power outage, how long can a power outage last during a major hurricane?**

   The response generated by ChatGPT:

   The length of a power outage during a major hurricane can vary greatly depending on the severity of the storm, the extent of the damage to the power grid, and the resources available to repair the damage. In some cases, power outages may only last a few hours, while in other cases they may last for several days or even weeks.

During a major hurricane, the damage to the power grid can be extensive, with downed power lines, damaged transformers, and flooded power stations. Repair crews must often wait until the storm has passed before they can begin repairs, which can further delay the restoration of power.

In some cases, utilities may also have to prioritize restoring power to critical infrastructure, such as hospitals and emergency services, before restoring power to individual homes and businesses.

If you are in an area that is affected by a hurricane and a power outage, it's important to be prepared for the possibility of an extended outage. Be sure to have emergency supplies on hand, such as flashlights, batteries, and non-perishable food. You may also want to consider purchasing a generator or alternative power source to help keep essential appliances running.

It's also important to stay informed about the latest updates on the power outage and restoration efforts from your local utility or emergency management agencies, and to follow any instructions or guidance provided by these organizations.

**Question 15. I rely on powered medical equipment and there is a power outage during a hurricane. What should I do?**

The response generated by ChatGPT:

If you rely on powered medical equipment and there is a power outage during a hurricane, it's important to take steps to ensure your safety and well-being. Here are some things you can do:

1. Notify your utility company: If you have not already done so, contact your utility company to let them know that you rely on powered medical equipment and that you are without power. They may be able to give you an estimated time for when the power will be restored     or provide you with information on nearby shelters or other resources.

2. Use backup power: If you have a backup generator, you can use it to power your medical equipment. Make sure the generator is located in a well-ventilated area, away from your home or any other buildings, and follow all safety instructions.

3. Contact your medical provider: If you are unable to use your medical equipment due to the power outage, contact your medical provider or healthcare facility for guidance. They may be able to provide you with alternative equipment or other solutions.

4. Seek shelter: If the power outage is expected to last for an extended period of time and you are unable to use your medical equipment, consider seeking shelter at a hospital or other medical facility. Contact your local emergency management agency or hospital in advance to arrange for accommodations.

5. Be prepared: It's important to be prepared for power outages by having a plan in place and emergency supplies on hand. Make sure you have backup batteries for your medical equipment, as well as flashlights, batteries, and non-perishable food. It may also be helpful to have a portable cell phone charger or other means of communication.

**Question 16. Our area is affected by a hurricane and power outage, how can I stay informed about power outages during a hurricane?**

The response generated by ChatGPT:

During a hurricane and power outage, it's important to stay informed about the latest updates on power outages and restoration efforts in your area. Here are some ways you can stay informed:

1. Sign up for alerts: Many utility companies offer alerts that you can sign up for to receive notifications about power outages in your area. These alerts can be sent via email, text message, or phone call.

2. Check your utility company's website: Your utility company s website may have a map or tool that shows the location and status of power outages in your area. Check the website periodically for updates.

3. Use social media: Many utility companies also use social media to provide updates on power outages and restoration efforts. Follow your utility company on social media platforms such as Twitter, Facebook, or Instagram for the latest updates.

4. Listen to local news: Tune in to local news stations on the radio or TV for the latest updates on power outages and restoration efforts in your area. 5. Use smartphone apps: There are several smartphone apps that provide information on power outages and restoration efforts. These apps may also allow you to report an outage and track the status of restoration efforts.

It' s important to have multiple ways of staying informed about power outages and restoration efforts during a hurricane and power outage, as communication systems may be disrupted. Be sure to also have a battery- powered or hand-cranked radio on hand for updates in case other communication methods are unavailable.

**Question 17. There is a boil water notice in our area because of flooding, what should I do?**

The response generated by ChatGPT:

If there is a boil water notice in your area due to flooding, it means that the water in your area may be contaminated and could cause illness if consumed. Here are some steps you can take to ensure the safety of your drinking water:

1. Boil water before use: Boil tap water for at least one minute before using it for drinking, cooking, brushing teeth, making ice, or washing dishes. This will help kill any bacteria or other harmful organisms that may be present in the water.

2. Use bottled water: If you have access to bottled water, use it for drinking and cooking instead of tap water.

3. Follow other recommendations from your local water authority: The specific instructions for a boil water notice may vary depending on the situation in your area. Check with your local water authority for specific instructions, such as whether it is safe to shower or bathe in the tap water, or if other precautions are necessary.

4. Use water filters: If you have a water filtration system, check to make sure it is designed to remove bacteria and other contaminants. Use only filters that are labeled as "NSF 53" or "NSF 58" for microbiological reduction.

5. Be cautious with water sources: During a flood, water sources such as wells and springs can become contaminated. Do not drink water from these sources unless you have tested it or treated it to ensure it is safe.

It's important to take a boil water notice seriously and follow the recommendations from your local water authority until the notice is lifted. Drinking contaminated water can lead to serious illness, especially for people with weakened immune systems, so it's important to take steps to ensure the safety of your drinking water.

**Question 18. My area is affected by flooding and roads are inundated, what should I do if I need medical care?**

The response generated by ChatGPT:

If your area is affected by flooding and the roads are inundated, it may be challenging to access medical care. Here are some steps you can take if you need medical assistance:

1. Call 911 if it is an emergency: If you have a medical emergency, such as chest pain, shortness of breath, or severe bleeding, call 911 immediately. The emergency responders will have the equipment and expertise to navigate through flooded areas and provide the necessary medical care.

2. Contact your healthcare provider: If you have a non-emergency medical need, such as a prescription refill or a routine check-up, contact your healthcare provider. They may be able to offer advice or provide a telemedicine appointment to address your medical needs.

3. Check with local authorities: In some cases, emergency medical services or medical clinics may be set up in affected areas to provide medical care. Check with your local authorities or emergency management services to see if any such services are available in your area.

4. Use alternative transportation: If you can access medical care but the roads are inundated, consider using alternative transportation, such as a boat, if it is safe to do so. However, it's important to be cautious and avoid unnecessary risks.

5. Evacuate if necessary: If your medical needs cannot be met in your area, consider evacuating to a location with better access to medical care. Be sure to bring any necessary medical supplies and medications with you.

Remember to take all necessary precautions and follow any advice or recommendations from local authorities during flooding. If you have a medical emergency, call 911 immediately, and if you have any concerns about your medical needs, contact your healthcare provider for guidance.